\documentclass[conference]{IEEEtran}
\usepackage{amsmath,amssymb,amsfonts}
\usepackage{graphicx}
\usepackage{subcaption}
\usepackage{textcomp}
\usepackage{xcolor}
\usepackage{booktabs}
\usepackage{svg}
\usepackage{fvextra}
\usepackage{multirow}
\usepackage{rotating}
\usepackage{adjustbox}
\usepackage[linesnumbered, ruled, vlined]{algorithm2e}
\usepackage{cite} %
\newcommand{\citet}[1]{\cite{#1}}
\newcommand{\citep}[1]{\cite{#1}}
\usepackage{url}

\DefineVerbatimEnvironment{CodeBlock}{Verbatim}{breaklines=true, breaksymbolleft=}

\def\BibTeX{{\rm B\kern-.05em{\sc i\kern-.025em b}\kern-.08em
    T\kern-.1667em\lower.7ex\hbox{E}\kern-.125emX}}

\begin{document}

\makeatletter
\newcommand{\linebreakand}{%
  \end{@IEEEauthorhalign}
  \hfill\mbox{}\par
  \mbox{}\hfill\begin{@IEEEauthorhalign}
}
\makeatother

\title{On the Difficulty of Selecting Few-Shot Examples for Effective LLM-based Vulnerability Detection}

\author{
\IEEEauthorblockN{Md Abdul Hannan\IEEEauthorrefmark{1}}
\IEEEauthorblockA{
Colorado State University\\
ma.hannan@colostate.edu
}\\
\IEEEauthorblockN{Limin Jia}
\IEEEauthorblockA{
Carnegie Mellon University\\
liminjia@andrew.cmu.edu
}
\and
\IEEEauthorblockN{Ronghao Ni\IEEEauthorrefmark{1}}
\IEEEauthorblockA{
Carnegie Mellon University\\
ronghaon@andrew.cmu.edu
}\\
\IEEEauthorblockN{Ravi Mangal}
\IEEEauthorblockA{
Colorado State University\\
ravi.mangal@colostate.edu
}
\and
\IEEEauthorblockN{Chi Zhang}
\IEEEauthorblockA{
Carnegie Mellon University\\
chiz5@andrew.cmu.edu
}\\
\IEEEauthorblockN{Corina S. Pasareanu}
\IEEEauthorblockA{
Carnegie Mellon University\\
pcorina@andrew.cmu.edu
}
}

\IEEEoverridecommandlockouts
\makeatletter\def\@IEEEpubidpullup{6.5\baselineskip}\makeatother
\IEEEpubid{\parbox{\columnwidth}{
		{\fontsize{7.5}{7.5}\selectfont Workshop on LLM Assisted Security and Trust Exploration (LAST-X) 2026 \\
			27 February 2026, San Diego, CA, USA \\
			ISBN 978-1-970672-05-3 \\
			https://dx.doi.org/10.14722/last-x.2026.23025  \\
			www.ndss-symposium.org}
}
\hspace{\columnsep}\makebox[\columnwidth]{}}

\maketitle

\begingroup\renewcommand\thefootnote{\IEEEauthorrefmark{1}}
\footnotetext{These authors contributed equally to this work.}
\endgroup

\begin{abstract}
Large language models (LLMs) have demonstrated impressive capabilities across a wide range of coding tasks, including summarization, translation, completion, and code generation. Despite these advances, detecting code vulnerabilities remains a challenging problem for LLMs. In-context learning (ICL) has emerged as an effective mechanism for improving model performance by providing a small number of labeled examples within the prompt. Prior work has shown, however, that the effectiveness of ICL depends critically on how these few-shot examples are selected.
In this paper, we study two intuitive criteria for selecting few-shot examples for ICL in the context of code vulnerability detection. The first criterion leverages model behavior by prioritizing samples on which the LLM consistently makes mistakes, motivated by the intuition that such samples can expose and correct systematic model weaknesses. The second criterion selects examples based on semantic similarity to the query program, using k-nearest-neighbor retrieval to identify relevant contexts.

We conduct extensive evaluations using open-source LLMs and datasets spanning multiple programming languages. Our results show that for Python and JavaScript, careful selection of few-shot examples can lead to measurable performance improvements in vulnerability detection. In contrast, for C and C++ programs, few-shot example selection has limited impact, suggesting that more powerful but also more expensive approaches, such as re-training or fine-tuning, may be required to substantially improve model performance.

\end{abstract}

\begin{IEEEkeywords}
LLMs, Few-shot examples, Code vulnerability detection
\end{IEEEkeywords}

\section{Introduction}

Large language models (LLMs) have demonstrated impressive capabilities for coding tasks including detection of software defects and vulnerabilities (such as resource leaks, use-after-free of dynamic memory during program operation, and denial-of-service attacks), variable misuse detection (i.e., programmers used wrong variables), code summarization (i.e., represent a code segment with a single word), and code generation and code completion~\cite{jiang2024survey,zhou2025large,11029737,husein2025large,wan2024deep}. In this paper we explore the applicability of LLMs for vulnerability detection. This is a particularly important application, as software vulnerabilities are prevalent in many software systems, posing serious risks such as compromising sensitive data and system failures. There are limited studies in previous work that show that state-of-the-art LLMs with few-shot-learning capabilities can achieve competitive results in detecting software vulnerabilities 
compared to previous machine learning techniques~\cite{zheng2024few,lin2025large}. 
However, the performance (measured in terms of precision, recall, F1 score) of these models remains low~\cite{ding2024vulnerability,khare2025understanding} preventing the deployment of these models in realistic settings.

In this paper, we explore black-box, prompt-based methods for improving the performance of LLMs on vulnerability detection. We seek to evaluate techniques for improving the efficacy of in-context learning (ICL)---the well-known phenomenon of LLM's exhibiting improved performance when the provided prompt includes a set of examples (referred to as the \emph{few-shot set}) that demonstrate the task to be performed---for the purpose of vulnerability detection. Previous works indicate that the impact of ICL on model performance is highly sensitive to the specific examples chosen 
~\cite{liu2021makes,min2022noisy,xu2023expertprompting}. For vulnerability detection, the few-shot set comprises programs along with the ground-truth label indicating whether they have a vulnerability or not.

We explore methods for choosing the example programs that comprise the few-shot set used in the code vulnerability detection task. In particular, we explore two natural algorithms
for choosing the few-shot examples. The first algorithm, \textbf{Learn-from-Mistakes} (LFM), is based on the idea that LLM performance on a sample is informative about its usefulness as a few-shot example. Therefore, LFM first queries the model (multiple times) on a potential example and records if the model is consistently mistaken (or correct) on the example. An example is added to the few-shot set only if the model consistently {\em makes mistakes on it}. The intuition is that the model does not know well how to reason about that example therefore adding it in the few-shot set will likely help rectify the model behavior, while a randomly chosen example may not be as useful. We also explore alternate versions of LFM where an example is added only if the model is consistently correct on it. The intuition here is that adding examples where the model is consistently correct can help reinforce the desired behavior and improve performance. The second algorithm, \textbf{Learn-from-Nearest-Neighbors} (LFNN), is based on the intuition that the semantic similarity between an example and the program under test can be a helpful guide in deciding if the example should be added to the few-shot set. In particular, LFNN adds the nearest neighbors to the set.
To compute semantic similarity, LFNN relies on code embedding models that map programs to embedding vectors. Similarity of vectors can be computed using metrics such as cosine similarity. 

We further explore different ways for combining LFM and LFNN to yield few-shot sets that are both semantically similar to the program under test and provide a context summarizing the model's past mistakes. Each combination yields a unique algorithm to construct the few-shot set for the vulnerability detection task and represents a specific trade-off between the two criteria---model performance and semantic similarity---for selecting examples. 

We evaluate these methods using two popular open-source models that are known to perform well on coding tasks, namely \textbf{Qwen2.5-Coder-7B-Instruct} and \textbf{Gemma-3-4B-it}, as well as a closed-source model, \textbf{GPT-5-mini}. We use four challenging datasets for the evaluation that include programs from a variety of languages such as C/C++, Python, and JavaScript. 

Our experimental results show that while certain standalone strategies, particularly LFNN, can improve baseline vulnerability detection performance, their effectiveness varies across models and datasets. LFM introduces a strong inductive bias that tends to skew predictions toward the positive class, which often limits its applicability. In contrast, the combined methods provide more stable and generally more balanced performance, although their benefits are not uniform. Depending on the underlying model and dataset, these combinations may yield meaningful gains in both accuracy and F1-score, or only modest or no improvements, reflecting the heterogeneous nature of the vulnerability detection task.

\section{Related Work}

\paragraph{\bf Vulnerability Detection}

Vulnerability detection is a long-standing challenge in software security. %
Over the years, a wide range of techniques have been developed to detect vulnerabilities, ranging from static program analysis~\cite{lipp2022empirical,marques2025automated} to dynamic testing and fuzzing~\cite{fuzzingSurvey,cassel2023nodemedic, nodemedic-fine}, and from these traditional methods to more recent machine-learning-based approaches~\cite{li2025automated,vuldeepecker,linevd,melicher_lighweight_2021}.

Machine-learning-based approaches aim to detect vulnerabilities by learning patterns directly from data. Early work extracted hand-crafted features such as token frequencies, code complexity metrics, or dependency patterns, and trained classifiers to separate %
vulnerable from non-vulnerable code~\cite{nam2015clami,medeiros2020vulnerable}.
More recent methods use representation learning to automatically encode code or abstract representations of code into vector spaces that possibly capture both syntax and semantics, enabling models to recognize patterns that are difficult to design manually. These approaches have shown promising results~\cite{devign,ivdetect,linevd,steenhoek2024dataflow,yang2024security}, but their effectiveness is often constrained by dataset quality and the challenge of generalizing across diverse projects~\cite{croft2023data,yadav2024r+,li2025out,safdar2025data}.

\paragraph{\bf LLM-based Vulnerability Detection}

Recent attempts to apply large models (LM) to vulnerability detection generally fall into two broad approaches. The first line of work treats LMs as embedding models, extracting vector representations of code and then either training lightweight classifiers on top of these embeddings or fine tuning the LM itself for the downstream task~\cite{linevd,hanif2022vulberta,steenhoek2023empirical}. Many of these methods also incorporate a program’s abstract representations such as Abstract Syntax Trees (ASTs), Control Flow Graphs (CFGs), or data flow features to complement the raw code input~\cite{steenhoek2024dataflow,yang2024security,devign,sysevr}. Early research in this category has often focused on models like CodeBERT~\cite{feng2020codebert} and GraphCodeBERT~\cite{guo2020graphcodebert}, with more recent work exploring larger foundation models such as Qwen~\cite{bai2023qwen,team2024qwen2,qwen2.5,yang2025qwen3} and LLaMA~\cite{codellama,dubey2024llama,touvron2023llama}.

The second line of work leverages the generative capabilities of modern large language models (LLM) more directly. Instead of relying only on embeddings, these approaches test the ability of LLMs to reason about and classify vulnerabilities through code understanding and natural language generation. Some studies evaluate models in their pretrained form, while others fine tune the models to improve task performance~\cite{zhou2025large}.
Several recent efforts reflect this trend.
\citet{du2024generalization} introduce VulLLM, a multi-task LLM framework that integrates vulnerability interpretation and data augmentation to significantly improve code vulnerability detection.
\citet{farr2025expert} 
leverage few-shot prompting to enhance out-of-the-box LLMs for vulnerability detection. \citet{du2024vul} proposes knowledge-level retrieval-augmented generation (RAG) for code vulnerability detection and reports sizable gains over baselines.

Our work falls into this second category. Specifically, we focus on evaluating LLMs out of the box in a few-shot classification setting.

\paragraph{\bf In-Context Learning and Example Selection}

In-context learning (ICL)~\cite{brown2020language} enables a model to adapt to a new task by conditioning on a small number of input output examples provided in the prompt. These examples serve as implicit supervision, enabling the model to infer the task format and desired output style without expensive fine tuning. ICL has been widely studied in natural language and code related tasks, where it has been shown to substantially improve model performance over zero-shot prompting~\cite{brown2020language,ramesh2025review,wei2022chain,bansal2019learning}.

A central question in ICL is how to select the few examples that are most useful for the model. Selection strategies can range from simple random sampling to more sophisticated methods that consider similarity between the test input and candidate examples, prior evidence about LLM behavior, or task specific heuristics~\cite{liu2021makes,min2022noisy,jiang2020can,rubin2022learning,xu2023expertprompting}.
\citet{liu2021makes} retrieve examples that are semantically-similar to a test sample to formulate its corresponding prompt. \citet{rubin2022learning} train an efficient dense retriever to select training examples as prompts at test time.
\citet{xu2023expertprompting} employ in-context learning to create expert profiles that condition LLM responses.
In our setting, we investigate new strategies for selecting few-shot examples and study how they influence the performance of LLMs on vulnerability detection.

\paragraph{\bf Prompt Optimization}

Prompt optimization broadly refers to techniques that improve how tasks are presented to large language models so that they yield more reliable outputs~\cite{pryzant2023automatic,sabbatella2024prompt}. Beyond manual design, recent work has begun to explore automated prompt optimization. DSPy~\cite{khattab2023dspycompilingdeclarativelanguage} introduces a declarative framework that compiles language model pipelines into self-improving programs. GEPA~\cite{agrawal2025gepareflectivepromptevolution} uses reflective natural-language feedback and evolutionary search to iteratively refine prompts. Maestro~\cite{wang2025maestrojointgraph} extends this line of work to multi-agent settings by jointly optimizing node configurations and the structure of agent graphs to better mitigate structural failure modes. These advances connect closely to the problem of example selection in ICL, as the choice and arrangement of input–output demonstrations influences model performance~\cite{lu2022fantastically,guo2024makes,agarwal2024many,bhope2025optiseq}. In our work, we treat the construction of the few-shot example set as a form of prompt optimization and study how different selection strategies affect the performance of LLMs for vulnerability detection.

\section{Two Algorithms: LFM and LFNN}
We explore two algorithms for choosing the examples presented in the LLM's context when performing vulnerability detection. The output of both these algorithms is a set of few-shot examples where each example is a program with a yes/no label indicating if a vulnerability is present or not.

\subsection{Learn-from-Mistakes (LFM)}

\SetKwInput{KwInput}{Input} 
\SetKwInput{KwOutput}{Output} 
\SetKwRepeat{Do}{do}{while}
\SetKwIF{If}{ElseIf}{Else}{if}{then}{else if}{else}{}
\begin{algorithm}[t]
\SetAlgoLined
\captionsetup{format=plain,labelfont=bf,justification=centering}
\caption{Learn from Mistakes (LFM)}
\label{alg:learn_from_mistake}
\KwInput{(1) Vulnerability detection model $f_{VD}$;
(2) Training dataset $D$ with $m$ labeled samples;
(3) Initial few-shot set $\mathcal{S}_{init}$ with $r$ labeled samples;
(4) Output few-shot set size $n$; (5) Boolean $st$ indicating stacked or unstacked version; (6) Number of iterations $k$; (7) Option $opt$ choosing between incorrect ($I$), correct ($C$), and gray ($G$)}
\KwOutput{Few-shot set $\mathcal{S}$}
\tcp{Default values of parameters are $st=\text{TRUE},k=1,opt=I$}
$\mathcal{S}_{C} \gets D,~\mathcal{S}_{I} \gets D$ \;
\ForEach{$idx \in \{1,\ldots,k\}$}{
$\mathcal{S}_{ctxt} \gets \mathcal{S}_{init},~\mathcal{S}_{Ci} \gets \emptyset, ~\mathcal{S}_{Ii} \gets \emptyset$\;
\ForEach{$(x, y) \in D \setminus \mathcal{S}_{init}$}{
Compute prediction $\hat{y} = f_{VD}(\mathcal{S}_{ctxt}; x)$\;
    \uIf{$\hat{y} \neq y$}{
        $\mathcal{S}_{Ii} \gets \mathcal{S}_{Ii} \cup \{(x, y)\}$\;
        \lIf{$st$}{$\mathcal{S}_{ctxt} \gets \mathcal{S}_{ctxt} \cup \{(x, y)\}$}
    }\Else{
        $\mathcal{S}_{Ci} \gets \mathcal{S}_{Ci} \cup \{(x, y)\}$\;
    }
}
$\mathcal{S}_C \gets \mathcal{S}_C \cap \mathcal{S}_{Ci}$\;
$\mathcal{S}_I \gets \mathcal{S}_I \cap \mathcal{S}_{Ii}$\;
}
$\mathcal{S}_G \gets D \setminus (\mathcal{S}_C \cup \mathcal{S}_I)$\;
$\mathcal{S}_{opt} \gets \mathcal{S}_I$;

\If{$opt = C$}{$\mathcal{S}_{opt} \gets \mathcal{S}_C$}
\ElseIf{$opt = G$}{$\mathcal{S}_{opt} \gets \mathcal{S}_G$}

\uIf{$st$}{
$\mathcal{S} \gets \mathcal{S}_{init}$\;
$\mathcal{S}_{rand} \gets \text{Uniformly draw } (n-|\mathcal{S}|) \text{ samples from } S_{opt} \setminus \mathcal{S}_{init} $\;
$\mathcal{S} \gets \mathcal{S} \cup \mathcal{S}_{rand}$\;
}\Else{
    $\mathcal{S} \gets \text{Uniformly draw } n \text{ samples from } S_{opt}$\;
}
\textbf{return} $\mathcal{S}$\;
\end{algorithm}

Algorithm~\ref{alg:learn_from_mistake} chooses few-shot examples based on the intuition that the correctness of the LLM response on an example is informative about its usefulness as a few-shot example. The algorithm makes a linear scan over a labeled dataset---for each sample in the dataset, it queries an LLM asking it to predict if the sample program has a vulnerability or not. This information is used to construct a few-shot set from this dataset (which we refer to as the \emph{training} dataset). In its default version, the algorithm, which we call \textbf{Learn-from-Mistakes} (LFM), operates under the assumption that the examples on which the LLM makes mistakes are more informative than random examples and should be added to the few-shot set. It iteratively updates the few-shot set based on incorrect predictions. However, the algorithm can be configured to run under various other settings that we describe next. Although some variants may also select correct or gray examples, we hold onto the name Learn-from-Mistakes because the selection process is defined through mistake-based scanning. Mistakes serve as the primary signal for identifying informative examples.

The primary inputs to the algorithm are the LLM $f_{VD}$, a labeled dataset $D$ with $m$ pairs of programs and their corresponding yes/no label indicating the presence or absence of a vulnerability, an initial set $\mathcal{S}_{init}$ of few-shot examples, and the desired number of examples $n$ in the few-shot set returned by the algorithm. Note that the set $\mathcal{S}_{init}$ can be empty. The remaining inputs to the algorithm are used to configure it. The boolean input $st$, which stands for \emph{stacked}, indicates whether the context used while querying the LLM during a run of the LFM algorithm should be iteratively refined or not. The input $k$ dictates the number of linear scans over the dataset. Multiple scans help deal with the non-determinism of LLM responses. Finally, the input $opt$ controls if the examples that are added to the few-shot set are the ones where the model makes a mistake or the ones where it is correct.

The algorithm begins by initializing sets $\mathcal{S}_C$ and $\mathcal{S}_I$ with the entire training dataset $D$ (line 1). The $\mathcal{S}_C$ tracks the examples from $D$ where the LLM is correct while $\mathcal{S}_I$ tracks the examples where it is incorrect. The algorithm then enters a loop (lines 2-14) and makes a linear scan over the dataset in each loop iteration. Before starting a scan, the algorithm initializes the few-shot set $\mathcal{S}_{ctxt}$ that will be used when querying the LLM during the linear scan. It also initializes sets $\mathcal{S}_{Ci}$ and $\mathcal{S}_{Ii}$ that track the correctly and incorrectly labeled examples during each iteration. In each linear scan (lines 4-11), the LLM $f_{VD}$ is queried on each of the examples in the dataset (except the ones in the initial set $\mathcal{S}_{init}$). Depending on whether the prediction is correct or not, the example is added to the set $\mathcal{S}_{Ci}$ (line 10) or $\mathcal{S}_{Ii}$ (line 7), respectively. Moreover, if the boolean input $st$ is set to TRUE, the few-shot set $\mathcal{S}_{ctxt}$ used to query the LLM during the current linear scan is updated whenever the model makes a mistake (line 8).
At the end of each iteration, the sets $\mathcal{S}_C$ and $\mathcal{S}_I$ of correctly and incorrectly labeled examples are updated (lines 12-13). Note that these sets track the examples that are \emph{consistently} labeled correctly of incorrectly by the LLM across the different iterations. This is enforced by the set intersection operations in lines 12 and 13. After all the linear scans are over and a final version of the sets $\mathcal{S}_C$ and $\mathcal{S}_I$ have been constructed, the algorithm also constructs a set $\mathcal{S}_G$ of examples where the LLM is not consistently correct or incorrect ($G$ stands for gray).

Finally, the few-shot set $\mathcal{S}$ to be returned by the algorithm is computed (lines 21-26). If the flag $st$ is set to TRUE, the examples in the initial set $\mathcal{S}_{init}$ are included in the set $\mathcal{S}$. The remaining samples in the set $\mathcal{S}$ are chosen from the appropriate sets $\mathcal{S}_C$, $\mathcal{S}_I$, and $\mathcal{S}_G$ (denoted by $\mathcal{S}_{opt}$) as dictated by the value of the $opt$ input (lines 23 and 26).

\subsection{Learn-from-Nearest-Neighbors (LFNN)}
\SetKwInput{KwInput}{Input} 
\SetKwInput{KwOutput}{Output} 
\SetKwRepeat{Do}{do}{while}
\SetKwIF{If}{ElseIf}{Else}{if}{then}{else if}{else}{}
\begin{algorithm}[t]
\SetAlgoLined
\captionsetup{format=plain,labelfont=bf,justification=centering}
\caption{Learn from Nearest Neighbors (LFNN)}
\label{alg:learn_from_nearest_neighbor_no_demo}
\KwInput{(1) Training dataset $D$ with $m$ labeled samples; (2) Query instance $x$; (3) Number of nearest neighbors $n$; (4) Encoder model $enc$}
\KwOutput{Nearest neighbor set $\mathcal{NN}_x$ for the given query instance $x$}
\tcp{Part 1: General pre-computation}
$K \gets \emptyset$\;
\ForEach{$(x_i, y_i) \in D$}{
    $K \gets K \cup (i,enc(x_i))$\;
}

\tcp{Part 2: Instance-specific computation}
$q \gets enc(x),~ C \gets \emptyset$\;
\ForEach{$(i,k) \in K$}{
    $C \gets C \cup (i,cosine(k,q))$\;
}
\tcp{$Top_n^2(C)$ returns the $n$ pairs from $C$ with the largest second components}
$\mathcal{NN}_x \gets \{D[i] ~|~(i,k) \in Top_n^2(C)\}$\;
$\textbf{return } \mathcal{NN}_x$\; 
\end{algorithm}

Algorithm~\ref{alg:learn_from_nearest_neighbor_no_demo}, which we call the \textbf{Learn-from-Nearest-Neighbors} (LFNN) algorithm, chooses few-shot examples based on the intuition that the program samples most \emph{similar} to the program under query are the most helpful in improving LLM performance. Although similar ideas have been proposed in the context of other applications~\cite{xu2023k}, in this work, we use this idea to improve the performance of LLMs for the vulnerability detection task. 

The inputs to the LFNN algorithm are a labeled dataset $D$ with $m$ pairs of programs and their corresponding yes/no label indicating the presence or absence of a vulnerability, the program $x$ under query, the number $n$ of nearest neighbors of $x$ (i.e., the size of the few-shot set $\mathcal{S}$) returned by the algorithm, and the encoder model $enc$ to be used. $enc$ is used to compute embedding vectors for programs which are then used for the nearest neighbor computation.

The algorithm begins with a query-agnostic phase. First, the set of embedding vectors $K$ is initialized to be an empty set (line 1). The algorithm then makes a pass over the dataset $D$, computes the embedding vector corresponding to each sample program in the dataset, and stores the pair of program index $i$ and the corresponding embedding vector in $K$ (lines 2-4). This part is independent of the query \( x \) and can be computed in advance. The key vectors are stored for later use with any query instance \( x \).

The next phase of the algorithm is query-specific. Given a query instance \( x \), the algorithm first computes the embedding vector $q$ for $x$ (line 5). It also initializes a set $C$ to record the similarity between $x$ and the programs in the dataset $D$. For each vector $k$ in the set of embedding vectors $K$, the algorithm calculates the cosine similarity between the embedding of $x$ and $k$ (lines 6-8), which measures the similarity between the query embedding and each program embedding from the dataset. The set $C$ records the indices of $k$ and the corresponding cosine similarity with $q$.

Finally, the algorithm selects the $n$ programs from $D$ that have the highest cosine similarities, identifying the most relevant neighbors to the query (Line 9). 
These selected programs form the nearest neighbor set $\mathcal{NN}_x$, which serves as the output of this algorithm.

Note that the general pre-computation phase of the algorithm need not be run for each query. Instead, the embedding vectors can be stored and then reused for each new query.

\section{Combining the Two Algorithms}
\label{sec:methods}

\SetKwInput{KwInput}{Input} 
\SetKwInput{KwOutput}{Output}
\SetKwProg{Def}{def}{:}{}
\SetKwRepeat{Do}{do}{while}
\SetKwIF{If}{ElseIf}{Else}{if}{then}{else if}{else}{}
\begin{algorithm}[t]
\SetAlgoLined
\captionsetup{format=plain,labelfont=bf,justification=centering}
\caption{Combined Methods}
\label{alg:method_1}
\KwInput{(1) Vulnerability detection model $f_{VD}$;
(2) Training dataset $D$ with $m$ labeled samples;
(3) Initial few-shot set $\mathcal{S}_{init}$ with $r$ labeled samples;
(4) LFM output few-shot set sizes $n_1$ and $n_2$; (5) Boolean $st$ indicating stacked or unstacked version; (6) Number of iterations $k$; (7) Option $opt$ choosing between incorrect ($I$), correct ($C$), and gray ($G$); (8) Query instance $x$; (9) Number of nearest neighbors $n_3$; (10) Encoder model $enc$}
\KwOutput{Few-shot set $\mathcal{S}$}
\Def{$method_1$()}{
    $\mathcal{S}_{LFM} \gets LFM(f_{VD},D,\mathcal{S}_{init},n_1,st,k,opt)$\;
    $\mathcal{S}_{LFNN} \gets LFNN(D,x,n_3,enc)$\;
    $\textbf{return } \mathcal{S}_{LFM} \cup \mathcal{S}_{LFNN}$\;
}
\BlankLine
\Def{$method_2$()}{
    $\mathcal{S}_{LFNN} \gets LFNN(D,x,n_3,enc)$\;
    $\mathcal{S}_{LFM} \gets LFM(f_{VD},D,\mathcal{S}_{LFNN},n_1,st,k,opt)$\;
    $\textbf{return } \mathcal{S}_{LFM}$\;
}
\BlankLine
\Def{$method_3$()}{
    $\mathcal{S}_{LFM} \gets LFM(f_{VD},D,\mathcal{S}_{init},n_2,st,k,opt)$\;
    $\mathcal{S}_{LFNN} \gets LFNN(D,x,n_3,enc)$\;
    $\textbf{return } LFM(f_{VD},\mathcal{S}_{LFNN},\mathcal{S}_{LFM},n_1,st,k,opt)$
}
\end{algorithm}

We explore three different strategies for combining LFM and LFNN to enhance the model's performance in detecting code vulnerabilities (Algorithm~\ref{alg:method_1}). The output of each of these combinations is a query-specific few-shot set of examples that is then used to predict the label for the query. Note that while there can be other ways of combining LFM and LFNN, we believe the three combinations we explore in this work represent the most natural starting point.

\paragraph{\textbf{Method 1}}
\label{subsec:approach-1}
In this method (lines 1-4), we combine the few-shot set from the LFM with the nearest neighbors of the query instance $x$  computed by LFNN. 
This method is the most straightforward and cost-effective compared to the subsequent two approaches.
We begin by constructing a few-shot set $\mathcal{S}_{LFM}$ with a total of $n_1$ samples. Note that this set is agnostic of the query $x$, so it just needs to be computed once for all the queries. Next, for each query instance $x$, we generate a unique few-shot set $\mathcal{S}_{LFNN}$ with the $n_3$ nearest neighbors of $x$. The final few-shot set is obtained by taking the union of the general few-shot set $\mathcal{S}_{LFM}$  with the nearest neighbors $\mathcal{S}_{LFNN}$ of $x$. Typically, in practice, the final few-shot set is composed of an equal number of samples from both the sources.

\paragraph{\textbf{Method 2}}
\label{subsec:approach-2}
In this method (lines 5-8), we use the nearest neighbors of the query instance $x$ as initial few-shot examples $\mathcal{S}_{init}$ for LFM. 
Our intuition is that compared to using no initial few-shot examples or using random examples with LFM, the use of nearest neighbors provides the model with starting knowledge that is closely related to the query instance $x$. As a result, the few-shot set constructed by LFM is specifically tailored to the program $x$ under query and therefore, can be more effective at improving the vulnerability detection capabilities of the model. 
Note that, in contrast to Method 1, we alter the order of applying the two algorithms such that both the calls (lines 6 and 7) generate distinct few-shot sets tailored for each query instance $x$. In other words, we are not able to reuse any computation across the different queries.
Although this approach is more resource-intensive, we hypothesize that this customized few-shot set could enhance model performance.

\paragraph{\textbf{Method 3}}
\label{subsec:approach-3}
In contrast to methods 1 and 2 that both invoke LFM and LFNN just once, method 3 (lines 9-12) invokes LFM twice. This method first invokes LFM (line 10) in a manner similar to method 1, generating a few-shot set $\mathcal{S}_{LFM}$ of size $n_2$. This first call to LFM is query-agnostic and therefore, only needs to be made once. Next, the method invokes LFNN (line 11), again in a manner similar to method 1 and generates a set of size $n_3$. As usual, the call to LFNN is query-specific and needs to be repeated for each query. Next, and unlike the other methods, LFM is invoked a second time. For this invocation, instead of using $D$ as the dataset, the few-shot set $\mathcal{S}_{LFNN}$ computed by LFNN is used as the dataset. This enables inclusion of only those examples in the final few-shot set that are most similar to the query while also accounting for the model correctness on these examples. Moreover, the second call to LFM uses the few-shot set $\mathcal{S}_{LFM}$ computed on line 10 as the initial set. The intuition here is that initializing LFM with these examples can make LFM aware of the examples on which the model makes a mistake (or is correct, depending on the $opt$ parameter) and thus, enable LFM to pick more effective examples for the final few-shot set. Note that the second call to LFM is also query-specific.

\section{Experiments}
\label{sec:expts}
In this section, we report on our experiments with the proposed methods, using open source models. We aim to answer the following research questions.

\paragraph{\bf Research Questions}
\begin{enumerate}
    \item 

    How do the proposed algorithms compare individually with baselines (zero-shot and few-shot settings) in helping large language models find vulnerabilities in code? 

    \item    How do different strategies for combining LFM and LFNN influence the overall performance of the model and how do they compare to using either strategy in isolation?
    
    \item 
    Are the performance improvements introduced by LFM and LFNN consistent across different large language models, or are they model-specific?

    \item 
    Does the programming language or other linguistic characteristics of the dataset influence the effectiveness of LFM, LFNN, and their combinations?

\end{enumerate}

\paragraph{\bf Datasets and Models}

For datasets, we consider established benchmarks such as PrimeVul, DiverseVul, SVEN \cite{ding2024vulnerability,DiverseVul,SVEN}. These are well-curated datasets, including pairs of code samples (vulnerable vs. non-vulnerable). We experiment with adding both vulnerable and non-vulnerable few-shot examples %
to better gauge the performance of the LLMs on the vulnerability detection task. We also leverage recent work on vulnerabilities in JavaScript programs~\cite{cassel2023nodemedic,nodemedic-fine} and obtained a copy of the dataset generated by their tools directly from the authors. We refer to this dataset as NodeMedic.

PrimeVul consists of a training set with 7578 samples, comprising 3789 pairs, and a test set with 870 samples. It has 112 unique CWEs. To mitigate computational overhead, the test set was downsampled to a representative subset of 200 examples. This sampling process was conducted based on the intersection of Common Weakness Enumerations (CWEs) found in both the training and test data. The final sampled test dataset is balanced, containing 100 vulnerable and non-vulnerable pairs, and 58 unique CWEs.

DiverseVul comprises  330492 unpaired samples, and it has 150 unique CWEs. To ensure computational tractability, the dataset was first partitioned into a primary training and a primary test set, following an 80:20 ratio which were unbalanced. Subsequently, based on the intersection of CWEs present in both splits, a final balanced sample was created. This resulted in a training set of 200 examples and a test set of 300 examples. Both of these sampled sets have 114 unique and common CWEs.

SVEN comprises 1440 training and 166 validation samples. This dataset was partitioned into two distinct subsets based on the programming language of the functions. The C/C++ subset, designated SVENC, consists of 756 training and 90 validation samples. The Python subset, SVENP, is composed of 684 training and 76 validation samples. SVENC and SVENP have 7 and 4 unique CWEs, respectively (both train and validation set).

NodeMedic dataset 
was provided by the authors of NodeMedic-FINE~\cite{nodemedic-fine}, a dynamic analysis tool that detects taint flows from package APIs to dangerous sinks that may enable arbitrary command injection or code execution. 
The dataset is divided into 1,506 training and 189 test samples, each corresponding to a Node.js package with potentially vulnerable dataflows reported by the tool. All reports are either automatically confirmed by NodeMedic-FINE~\cite{nodemedic-fine} or manually verified by its authors.

To facilitate semantic code retrieval, we employed a specialized encoder model from the Salesforce SFR family: "Salesforce/SFR-Embedding-Code-400M\_R" \cite{liu2024codexembed}. 
These vector embeddings are used in the nearest neighbor search, which utilizes cosine distance to identify the closest matches for any given query.

We conducted experiments on two open-source models, "Qwen-2.5-Coder-7B-Instruct" \cite{hui2024qwen2} and "gemma-3-4b-it" \cite{gemma_2025}, as well as the closed-source GPT-5-mini (the \textit{gpt-5-mini-2025-08-07} snapshot), to assess the generalizability of our proposed techniques. The Qwen model was selected for its strong, well-documented proficiency on coding tasks and very large context window, while the Gemma model was chosen for its computational efficiency and competitive performance given its parameter count. GPT-5-mini serves as a high-quality closed-source baseline to contextualize the behavior of the open models. The combination of these models allows evaluation across differing trade-offs of capability and resource requirements, while staying within our available computational budget.

\begin{table*}[t]
\caption{Results for all approaches (ZS = zero-shot, R-FS = random few-shot, LFM = Learn-from-Mistakes, LFNN = Learn-from-Nearest-Neighbors, CM = Combined Method) across three models. Gemma and Qwen results are mean over five runs; GPT results are based on one run.
A dash (‘–’) indicates that the metric is undefined in at least one of the runs due to division by zero.}
\label{table:main_no_std}
\centering
\begin{tabular}{llrrrrrrrrrrrr}
\toprule
Dataset & Approach & \multicolumn{4}{c}{Gemma-3-4b-it} & \multicolumn{4}{c}{GPT-5-mini} & \multicolumn{4}{c}{Qwen-2.5-Coder} \\
\cmidrule(lr){3-6} \cmidrule(lr){7-10} \cmidrule(lr){11-14}
& & Acc & Prec & Recall & F1 & Acc & Prec & Recall & F1 & Acc & Prec & Recall & F1 \\
\midrule
\multirow{7}{*}{\rotatebox{90}{DiverseVul}} & ZS & 0.619 & 0.770 & 0.340 & 0.472 & 0.593 & 0.588 & 0.627 & 0.607 & 0.497 & 0.000 & 0.000 & 0.000 \\
& R-FS & 0.545 & 0.571 & 0.467 & 0.480 & 0.560 & 0.598 & 0.367 & 0.455 & 0.599 & 0.741 & 0.303 & 0.411 \\
& LFM & 0.525 & 0.840 & 0.061 & 0.114 & 0.617 & 0.647 & 0.513 & 0.573 & 0.500 & 0.500 & 1.000 & 0.667 \\
& LFNN & 0.512 & 0.511 & 0.537 & 0.524 & 0.590 & 0.624 & 0.453 & 0.525 & 0.659 & 0.632 & 0.759 & 0.690 \\
& CM 1 & 0.548 & 0.530 & 0.868 & 0.658 & 0.587 & 0.610 & 0.480 & 0.537 & 0.638 & 0.606 & 0.792 & 0.686 \\
& CM 2 & 0.515 & 0.508 & 0.973 & 0.667 & 0.560 & 0.573 & 0.473 & 0.518 & 0.597 & 0.557 & 0.949 & 0.702 \\
& CM 3 & 0.499 & 0.499 & 0.985 & 0.663 & 0.570 & 0.585 & 0.480 & 0.527 & 0.531 & 0.517 & 0.988 & 0.678 \\
\midrule
\multirow{7}{*}{\rotatebox{90}{NodeMedic}} & ZS & 0.506 & 0.765 & 0.460 & 0.574 & 0.757 & 0.786 & 0.912 & 0.845 & 0.375 & 0.870 & 0.162 & 0.273 \\
& R-FS & 0.687 & 0.726 & 0.914 & 0.804 & 0.794 & 0.845 & 0.876 & 0.860 & 0.632 & 0.748 & 0.750 & 0.739 \\
& LFM & 0.725 & 0.725 & 1.000 & 0.840 & 0.767 & 0.789 & 0.927 & 0.852 & 0.720 & 0.723 & 0.993 & 0.837 \\
& LFNN & 0.758 & 0.759 & 0.975 & 0.854 & 0.788 & 0.849 & 0.861 & 0.855 & 0.713 & 0.768 & 0.866 & 0.814 \\
& CM 1 & 0.701 & 0.755 & 0.870 & 0.808 & 0.788 & 0.839 & 0.876 & 0.857 & 0.751 & 0.776 & 0.923 & 0.843 \\
& CM 2 & 0.565 & 0.719 & 0.657 & 0.686 & 0.794 & 0.840 & 0.883 & 0.861 & 0.735 & 0.748 & 0.958 & 0.840 \\
& CM 3 & 0.722 & 0.728 & 0.984 & 0.837 & 0.794 & 0.846 & 0.883 & 0.864 & 0.720 & 0.724 & 0.991 & 0.837 \\
\midrule
\multirow{7}{*}{\rotatebox{90}{PrimeVul}} & ZS & 0.599 & 0.723 & 0.358 & 0.472 & 0.535 & 0.524 & 0.750 & 0.617 & 0.500 & -- & 0.002 & 0.004 \\
& R-FS & 0.522 & 0.553 & 0.373 & 0.422 & 0.560 & 0.556 & 0.600 & 0.577 & 0.503 & 0.528 & 0.232 & 0.304 \\
& LFM & 0.516 & 0.726 & 0.437 & 0.336 & 0.550 & 0.557 & 0.490 & 0.521 & 0.500 & 0.500 & 1.000 & 0.667 \\
& LFNN & 0.527 & 0.520 & 0.648 & 0.571 & 0.555 & 0.549 & 0.620 & 0.582 & 0.483 & 0.390 & 0.060 & 0.104 \\
& CM 1 & 0.543 & 0.527 & 0.885 & 0.659 & 0.575 & 0.577 & 0.600 & 0.588 & 0.473 & 0.462 & 0.336 & 0.383 \\
& CM 2 & 0.516 & 0.508 & 0.964 & 0.666 & 0.555 & 0.549 & 0.610 & 0.578 & 0.455 & 0.445 & 0.370 & 0.404 \\
& CM 3 & 0.500 & 0.500 & 0.982 & 0.663 & 0.555 & 0.546 & 0.650 & 0.594 & 0.499 & 0.499 & 0.986 & 0.663 \\
\midrule
\multirow{7}{*}{\rotatebox{90}{SVENC}} & ZS & 0.478 & 0.474 & 0.400 & 0.434 & 0.533 & 0.523 & 0.756 & 0.618 & 0.500 & -- & 0.000 & 0.000 \\
& R-FS & 0.506 & 0.503 & 0.778 & 0.564 & 0.600 & 0.579 & 0.733 & 0.647 & 0.491 & -- & 0.422 & 0.345 \\
& LFM & 0.500 & 0.500 & 1.000 & 0.667 & 0.544 & 0.543 & 0.556 & 0.549 & 0.500 & 0.500 & 1.000 & 0.667 \\
& LFNN & 0.471 & 0.428 & 0.187 & 0.258 & 0.567 & 0.554 & 0.689 & 0.614 & 0.504 & 0.567 & 0.040 & 0.075 \\
& CM 1 & 0.520 & 0.513 & 0.813 & 0.629 & 0.578 & 0.566 & 0.667 & 0.612 & 0.520 & 0.527 & 0.387 & 0.442 \\
& CM 2 & 0.536 & 0.520 & 0.947 & 0.671 & 0.589 & 0.574 & 0.689 & 0.626 & 0.493 & 0.495 & 0.636 & 0.556 \\
& CM 3 & 0.500 & 0.500 & 0.938 & 0.652 & 0.611 & 0.596 & 0.689 & 0.639 & 0.496 & 0.498 & 0.982 & 0.661 \\
\midrule
\multirow{7}{*}{\rotatebox{90}{SVENP}} & ZS & 0.587 & 0.597 & 0.537 & 0.565 & 0.763 & 0.727 & 0.842 & 0.780 & 0.705 & 0.965 & 0.426 & 0.591 \\
& R-FS & 0.558 & 0.561 & 0.632 & 0.584 & 0.789 & 0.806 & 0.763 & 0.784 & 0.616 & 0.707 & 0.432 & 0.514 \\
& LFM & 0.500 & 0.500 & 0.905 & 0.644 & 0.776 & 0.744 & 0.842 & 0.790 & 0.500 & 0.500 & 1.000 & 0.667 \\
& LFNN & 0.634 & 0.641 & 0.611 & 0.625 & 0.895 & 0.941 & 0.842 & 0.889 & 0.792 & 0.845 & 0.716 & 0.775 \\
& CM 1 & 0.553 & 0.546 & 0.616 & 0.579 & 0.882 & 0.892 & 0.868 & 0.880 & 0.787 & 0.793 & 0.779 & 0.785 \\
& CM 2 & 0.632 & 0.639 & 0.605 & 0.622 & 0.895 & 0.917 & 0.868 & 0.892 & 0.697 & 0.660 & 0.816 & 0.729 \\
& CM 3 & 0.521 & 0.512 & 0.942 & 0.663 & 0.803 & 0.795 & 0.816 & 0.805 & 0.600 & 0.558 & 0.963 & 0.707 \\
\bottomrule
\end{tabular}
\end{table*}

\paragraph{\bf Experiment Setup}

For our baselines, we consider a \emph{zero-shot} setting, i.e., no examples in context, and a \emph{random few-shot} setting where twenty vulnerable and non-vulnerable examples are drawn at random, from our train datasets.
To address RQ1, we create several variants of the LFM algorithm with different parameter settings. However, since the performance differences are minimal, we only report the results for the following configurations: $n=20, st=\text{TRUE}, k=1, opt=I$ and an initial few-shot set $\mathcal{S}_{init}$ with an empty set.
We run the LFNN algorithm with parameter settings $n=20$. To answer RQ2, we run the combined methods with the following parameter settings (for all three combinations, the initial few-shot set $\mathcal{S}_{init}$ has 5 examples that are randomly drawn from the train dataset):
\begin{itemize}
    \item Combined Method 1: $n_1=10,st=\text{TRUE},k=1,opt=I,n_3=10$
    \item Combined Method 2: $n_1=20,st=\text{TRUE},k=1,opt=I,n_3=5$ 
    \item Combined Method 3: $n_1=10,n_2=15,st=\text{TRUE},k=1,opt=I,n_3=20$
\end{itemize}

\paragraph{\bf Prompt Used for Vulnerability Detection}
\label{sec:setup_llm}
The prompt shown here is used when running LFM as well as for evaluating the vulnerability detection capabilities of the LLM for the experiments reported in Section~\ref{sec:expts}.
To instruct the model clearly, we provided a concise and explicit system instruction, guiding the model to behave strictly as a security expert and to output responses in a standardized format. The exact system-level instruction used to prime the LLM is defined as follows:

\begin{CodeBlock}
You are a security expert that is good at static program analysis.
First, you will be given some examples of vulnerable and non-vulnerable codes indicated through Yes and No. There can be no examples too.You will be given a piece of code. Your task is to analyze whether 
it contains a security vulnerability.
Please only reply with one of the following options:
(1) YES: A security vulnerability detected.
(2) NO: No security vulnerability.

Only reply with one of the options above. Do not include any further information.
\end{CodeBlock}

For each code snippet \(x\), we first construct a few-shot set $\mathcal{S}$ containing representative examples of code snippets, each annotated as either \emph{vulnerable} or \emph{non-vulnerable}. These labeled examples serve as context to guide the model's decision by explicitly illustrating the desired behavior. The few-shot prompt we use is as follows (\texttt{System, User, Assistant} refer to the roles used for prompting the model):

\begin{CodeBlock}
System:
  (System Instruction from above)

User:
  Code: <Example 1: Vulnerable code snippet>
  Answer:

Assistant:
  YES

User:
  Code: <Example 2: Non-vulnerable code snippet>
  Answer:

Assistant:
  NO

... (remaining few-shot examples)

User:
  Code: {code}
  Answer:
\end{CodeBlock}

In our experiments, all dataset sampling is performed with a fixed random seed to ensure reproducibility.
For the Qwen and Gemma models, deterministic algorithms are enabled, and all random number generators in Python and PyTorch are explicitly seeded to ensure deterministic behavior under our hardware and software setup.
For GPT-5-Mini, we set the API seed parameter to a fixed integer to encourage deterministic generation. However, as documented by OpenAI\footnote{\url{https://cookbook.openai.com/examples/reproducible_outputs_with_the_seed_parameter}}, setting this seed does not guarantee consistent outputs because changes to the backend system, such as model version updates, can still affect the generations.

\paragraph{\bf Results}

We present a detailed analysis of the experimental results from our evaluation of three LLMs from the Gemma, GPT, and Qwen families across the four datasets (Results are presented separately for SVENC and SVENP).  All findings are summarized in Table~\ref{table:main_no_std}, which reports the mean performance metrics over five independent runs for the Gemma and Qwen models using five different random seeds. The standard deviations between runs are minimal, so they are not included in the table because of space constraints. For the GPT model, we report the results from a single run. Our analysis is structured on a per-dataset basis to highlight the varying effectiveness of each few-shot selection approach under different data distributions and model capabilities.

On the \textbf{NodeMedic} dataset, LFM and LFNN demonstrate substantial performance gains over the zero-shot (ZS) and random few-shot (R-FS) baselines across all models. For the Gemma model, LFNN achieves the highest F1-score of 0.854, a significant improvement from the ZS baseline of 0.574. The LFM approach pushed the model to a perfect recall of 1.000 and achieved an F1-score of 0.840. The Qwen model shows a similar trend, with LFM and LFNN improving the F1-score to 0.837 and 0.814, respectively. GPT-5-mini, a more powerful model, exhibits strong performance even with the R-FS baseline (F1-score of 0.860), but our adaptive methods still provide a slight edge. Overall, there was a good balance between accuracy and F1-score with combined method 1 and 3 , and LFNN (accuracy: 0.701, 0.722, 0.758 respectively).

The \textbf{SVENP} dataset also yields strong results, but with a key difference: the ZS baseline is notably more effective here, especially for Qwen (0.965 Precision) and GPT-5-mini (0.780 F1-score). This indicates that the patterns in SVENP align well with the models' pre-trained knowledge or maybe the models are in general better at analyzing python source code. Despite the strong baseline, LFNN improves the performance by achieving top-tier F1-scores of 0.889 (GPT) and 0.775 (Qwen). For the Gemma model, LFNN boosts the F1-score from 0.565 (ZS) to 0.625, and accuracy from 0.587 (ZS) to 0.634. A critical observation is the behavior of LFM with the Qwen model; it again defaults to predicting the positive class for all instances (1.000 Recall). This highlights LFM's tendency to act as a powerful bias amplifier, which is effective when correcting false negatives but can be overly simplistic and increase false positives. The Combined Methods (CM), particularly on GPT-5-mini, achieve the highest overall F1-scores (e.g., 0.892 for CM 2), demonstrating that integrating both mistake-based and similarity-based signals is optimal when the baseline performance is already high.

The \textbf{DiverseVul} dataset presents a more complex challenge. Here, the ZS baseline for Qwen completely fails, predicting the negative class for all samples and resulting in an F1-score of 0.000. In contrast, the Gemma ZS baseline is more reasonable, with an F1-score of 0.472. For the Qwen model, LFNN is the most effective individual strategy, raising the F1-score to 0.690. However, the most compelling finding on this dataset comes from the Gemma model. While LFM performs poorly (0.114 F1-score) and LFNN offers only a modest improvement (0.524 F1-score), the Combined Methods deliver the best performance. CM 1, CM 2, and CM 3 achieve F1-scores of 0.658, 0.667, and 0.663, with corresponding accuracies of 0.548, 0.515, and 0.499, respectively. The robustness of the Combined Methods indicates that a blended approach is necessary to navigate the diverse patterns present in the data.

The \textbf{PrimeVul} and \textbf{SVENC} datasets underscore the importance of F1-score over accuracy. On both datasets, the Qwen model's ZS baseline fails by uniformly predicting the negative class (0.000 F1-score), while the LFM approach predictably does the opposite, classifying all samples as positive (0.667 F1-score). In both cases, the accuracy is a misleading 0.500, masking these divergent failure modes. A crucial observation from these two datasets is the significant degradation of the LFNN method. For Qwen, LFNN yields very low F1-scores of 0.104 on PrimeVul and 0.075 on SVENC. This is a stark contrast to its success on NodeMedic and SVENP. This failure implies that for these datasets, the nearest neighbors examples are not helpful. Once again, the Combined Methods demonstrate greater resilience, particularly for the Gemma model. On SVENC, CM 2 elevates the F1-score to 0.671, the highest for that model. This pattern reinforces the conclusion that when simpler adaptive heuristics like LFNN fail, a more robust, multi-faceted example selection strategy is required to achieve better performance.

Finally, one interesting insight was hard examples (LFM) in context try to bias the model towards greater recall and lower precision. As CM3 applies LFM on LFNN examples keeping first LFM's result as context, it also keeps that trend mostly.

\paragraph{\bf Different Variants of Learn-from-Mistakes (LFM)}
\begin{table*}[!htbp]
\centering
\makebox[\textwidth][c]{
\rotatebox{90}{
\begin{minipage}{\dimexpr\textheight-1.5\baselineskip\relax}
\caption{Performance comparison of different LFM variations. Each cell shows the mean and standard deviation over five runs. For the LFM variation names, the following abbreviations are used: S = stacked, U = unstacked, 1× = one iteration, m× = more than one iteration, inc. = incorrect, corr. = correct, and gray = gray cases. ‘–’ indicates metrics that are undefined due to division by zero.
}
\label{table:rq5_combined}
\centering
\resizebox{!}{0.22\textheight}{
\begin{tabular}{llrrrrrrrr}
\toprule
Dataset & LFM Variations & \multicolumn{4}{c}{Gemma-3-4b-it} & \multicolumn{4}{c}{Qwen-2.5-Coder} \\
\cmidrule(lr){3-6} \cmidrule(lr){7-10}
 & & Accuracy & Precision & Recall & F1 Score & Accuracy & Precision & Recall & F1 Score \\
\midrule
\multirow{5}{*}{DiverseVul} & S-1× (inc.) & 0.525 (±0.003) & 0.840 (±0.056) & 0.061 (±0.005) & 0.114 (±0.009)  & 0.500 (±0.000) & 0.500 (±0.000) & 1.000 (±0.000) & 0.667 (±0.000)  \\
 & U-1× (inc.)  & 0.502 (±0.013) & 0.501 (±0.007) & 0.940 (±0.025) & 0.654 (±0.010)  & 0.502 (±0.016) & 0.501 (±0.010) & 0.844 (±0.063) & 0.628 (±0.022)  \\
 & U-m× (inc.) & 0.507 (±0.010) & 0.504 (±0.005) & 0.995 (±0.008) & 0.669 (±0.003)  & 0.527 (±0.039) & 0.516 (±0.025) & 0.976 (±0.045) & 0.674 (±0.009)  \\
 & U-m× (corr.) & 0.597 (±0.053) & 0.778 (±0.056) & 0.295 (±0.189) & 0.383 (±0.204)  & 0.621 (±0.065) & 0.752 (±0.027) & 0.365 (±0.199) & 0.455 (±0.199)  \\
 & U-m× (gray) & 0.513 (±0.000) & 0.667 (±0.000) & 0.053 (±0.000) & 0.099 (±0.000)  & 0.598 (±0.061) & 0.713 (±0.059) & 0.341 (±0.226) & 0.415 (±0.199)  \\
\midrule
\multirow{5}{*}{NodeMedic} & S-1× (inc.) & 0.725 (±0.000) & 0.725 (±0.000) & 1.000 (±0.000) & 0.840 (±0.000)  & 0.720 (±0.000) & 0.723 (±0.000) & 0.993 (±0.000) & 0.837 (±0.000)  \\
 & U-1× (inc.)  & 0.273 (±0.004) & 0.500 (±0.707) & 0.001 (±0.003) & 0.003 (±0.006)  & 0.274 (±0.005) & 0.333 (±0.422) & 0.004 (±0.006) & 0.009 (±0.011)  \\
 & U-m× (inc.) & 0.277 (±0.000) & - & 0.000 (±0.000) & 0.000 (±0.000)  & 0.272 (±0.003) & -- & 0.000 (±0.000) & 0.000 (±0.000)  \\
 & U-m× (corr.) & 0.725 (±0.000) & 0.725 (±0.000) & 1.000 (±0.000) & 0.840 (±0.000)  & 0.724 (±0.002) & 0.727 (±0.001) & 0.991 (±0.003) & 0.839 (±0.001)  \\
 & U-m× (gray) & 0.506 (±0.003) & 0.765 (±0.005) & 0.460 (±0.000) & 0.574 (±0.001)  & 0.467 (±0.100) & 0.748 (±0.022) & 0.412 (±0.245) & 0.493 (±0.158)  \\
\midrule
\multirow{5}{*}{PrimeVul} & S-1× (inc.) & 0.516 (±0.013) & 0.726 (±0.186) & 0.437 (±0.459) & 0.336 (±0.270) & 0.500 (±0.000) & 0.500 (±0.000) & 1.000 (±0.000) & 0.667 (±0.000)  \\
 & U-1× (inc.)  & 0.507 (±0.010) & 0.529 (±0.054) & 0.773 (±0.352) & 0.551 (±0.212) & 0.497 (±0.027) & 0.489 (±0.042) & 0.466 (±0.137) & 0.470 (±0.092)  \\
 & U-m× (inc.) & 0.496 (±0.009) & 0.397 (±0.221) & 0.034 (±0.022) & 0.061 (±0.038) & 0.501 (±0.005) & 0.502 (±0.005) & 0.864 (±0.219) & 0.621 (±0.077)  \\
 & U-m× (corr.) & 0.499 (±0.002) & 0.499 (±0.001) & 0.886 (±0.223) & 0.626 (±0.079) & 0.489 (±0.010) & 0.386 (±0.090) & 0.034 (±0.010) & 0.062 (±0.018)  \\
 & U-m× (gray) & 0.499 (±0.002) & 0.496 (±0.009) & 0.366 (±0.284) & 0.357 (±0.202) & 0.498 (±0.017) & 0.529 (±0.141) & 0.246 (±0.241) & 0.267 (±0.198)  \\
\midrule
\multirow{5}{*}{SVENC} & S-1× (inc.) & 0.500 (±0.000) & 0.500 (±0.000) & 1.000 (±0.000) & 0.667 (±0.000)  & 0.500 (±0.000) & 0.500 (±0.000) & 1.000 (±0.000) & 0.667 (±0.000)  \\
 & U-1× (inc.)  & 0.500 (±0.000) & - & 0.000 (±0.000) & 0.000 (±0.000)  & 0.500 (±0.000) & -- & 0.000 (±0.000) & 0.000 (±0.000)  \\
 & U-m× (inc.) & 0.500 (±0.000) & - & 0.000 (±0.000) & 0.000 (±0.000)  & 0.500 (±0.000) & -- & 0.000 (±0.000) & 0.000 (±0.000)  \\
 & U-m× (corr.) & 0.504 (±0.009) & 0.503 (±0.006) & 0.960 (±0.069) & 0.659 (±0.013)  & 0.493 (±0.009) & 0.496 (±0.005) & 0.911 (±0.056) & 0.642 (±0.016)  \\
 & U-m× (gray) & 0.478 (±0.000) & 0.474 (±0.000) & 0.400 (±0.000) & 0.434 (±0.000)  & 0.502 (±0.011) & -- & 0.080 (±0.109) & 0.114 (±0.142)  \\
\midrule
\multirow{5}{*}{SVENP} & S-1× (inc.) & 0.500 (±0.000) & 0.500 (±0.000) & 0.905 (±0.013) & 0.644 (±0.003)  & 0.500 (±0.000) & 0.500 (±0.000) & 1.000 (±0.000) & 0.667 (±0.000)  \\
 & U-1× (inc.)  & 0.558 (±0.067) & 0.571 (±0.083) & 0.600 (±0.127) & 0.573 (±0.050)  & 0.590 (±0.051) & 0.607 (±0.074) & 0.547 (±0.111) & 0.567 (±0.062)  \\
 & U-m× (inc.) & 0.540 (±0.033) & 0.567 (±0.079) & 0.447 (±0.130) & 0.483 (±0.074)  & 0.571 (±0.054) & 0.569 (±0.065) & 0.721 (±0.132) & 0.624 (±0.041)  \\
 & U-m× (corr.) & 0.524 (±0.023) & 0.522 (±0.018) & 0.621 (±0.154) & 0.557 (±0.063)  & 0.629 (±0.038) & 0.712 (±0.045) & 0.437 (±0.133) & 0.528 (±0.110)  \\
 & U-m× (gray) & 0.547 (±0.062) & 0.549 (±0.064) & 0.742 (±0.145) & 0.619 (±0.034)  & 0.626 (±0.038) & 0.767 (±0.120) & 0.437 (±0.175) & 0.519 (±0.106)  \\
\bottomrule
\end{tabular}
}
\end{minipage}
}
}
\end{table*}

To explore the impact of different LFM variants, we run LFM with the following different parameter settings:
\begin{itemize}
    \item 
    S-1×: $n=20,st=\text{TRUE},k=1,opt=I$
    \item 
    U-1×: $n=20,st=\text{FALSE},k=1,opt=I$
    \item 
    U-m× (inc.): $n=20,st=\text{FALSE},k=5,opt=I$
    \item 
    U-m× (corr.):$n=20,st=\text{FALSE},k=5,opt=C$
    \item 
    U-m× (gray):$n=20,st=\text{FALSE},k=5,opt=G$
\end{itemize}

\noindent For all the unstacked variations, the initial few-shot set $\mathcal{S}_{init}$ has 20 examples that are randomly drawn from the train dataset while $\mathcal{S}_{init}$ is the empty set for the stacked version.

Table~\ref{table:rq5_combined} shows a comprehensive breakdown of performance across different LFM configurations. It is evident that the method's effectiveness is highly sensitive to its parameterization and the specific characteristics of the dataset.

First, we have already stated that the default LFM method used in our main experiments, S-1x (inc.), consistently induces a strong bias towards positive predictions. This is most evident with the Qwen model, where it achieves a perfect 1.000 recall four times. However, this perfect recall on balanced datasets results in an uninformative accuracy of 0.500 and a misleadingly high F1-score of 0.667. For Gemma, this method shows highly divergent performance. It is effective on NodeMedic, achieving both high accuracy (0.725) and a high F1-score (0.840). Yet, on SVENC, it shows the biased trend: 0.667 F1-score with 0.500 accuracy. On DiverseVul and PrimeVul, it fails on both metrics, with F1-scores of 0.114 and 0.336, respectively. This confirms that it often trades all precision for recall.

Second, the comparison between stacked (S-1x) and unstacked (U-1x) methods highlights the critical impact of the initial prompt $\mathcal{S}_{init}$. For the NodeMedic and SVENC datasets, the U-1x (inc.) variant fails catastrophically. On NodeMedic, accuracy plummets to 0.273 for Gemma, and the F1-score becomes near-zero (0.003). On SVENC, it results in 0.000 recall, indicating a complete shift to negative-class bias. Conversely, on DiverseVul and PrimeVul with the Gemma model, where the stacked method failed, the unstacked version provides a better balance. On DiverseVul, it improves the F1-score from 0.114 to 0.654, though accuracy remains low at 0.502. This indicates that while the random examples help, they do not solve the accuracy/F1 trade-off, instead achieving a high F1 (0.654) through high recall (0.940) and low precision (0.501).

Third, a comparison of the unstacked, multi-iteration (U-mx) variants reveals dataset-dependent findings. For NodeMedic and SVENC, where learning from incorrect examples (opt=I) failed, learning from correct examples (opt=C) is highly effective. On NodeMedic, this U-mx (corr.) variant restores both high accuracy (0.725) and a high F1-score (0.840) for Gemma, mirroring the performance of the original stacked LFM. This demonstrates an ideal balance. On SVENC, however, the same method only restores the high F1-score (0.659) while accuracy remains low (0.504), indicating it learned to trade precision for high recall (0.960). For these same two datasets, the U-m$\times$ (gray) variant was largely ineffective. Gemma's performance regressed to F1-scores of $0.574$ (NodeMedic) and $0.434$ (SVENC), close to the ZS baseline.

This pattern is completely reversed on the DiverseVul dataset. Here, learning from incorrect examples (opt=I) yields a high F1-score (0.669 for Gemma) but poor accuracy (0.507). In contrast, learning from correct examples (opt=C) achieves a much better accuracy (0.597) at the cost of a poor F1-score (0.383), presenting a clear trade-off. The U-m$\times$ (gray) variant was similarly ineffective on this dataset, achieving a very low F1-score of $0.099$ for Gemma. The PrimeVul dataset presents the most significant finding: a direct contradiction between the models. Gemma achieves its best unstacked F1-score (0.626) with U-mx (corr.), though its accuracy remains low (0.499). It fails with U-mx (inc.) (0.061 F1). Qwen's performance is the exact inverse, performing best with U-mx (inc.) (F1 0.621, Acc 0.501) and failing with U-mx (corr.) (F1 0.062, Acc 0.489). Here, the U-m$\times$ (gray) variant for Gemma on PrimeVul landed in the middle, with a modest F1-score of $0.357$.

In summary, the LFM's behavior is not monolithic. The stacked, single-iteration approach is a predictable bias-inducer, sacrificing accuracy for recall. The unstacked, multi-iteration approaches are highly contingent on the type of examples used for learning (i.e., whether $opt=I$, $opt=C$, or $opt=G$), and no single strategy has been proven universally superior. Furthermore, our analysis highlights a persistent tension between optimizing for F1-score and accuracy. On balanced datasets, in some cases, LFM variants achieve better F1-score by inducing a strong recall bias, which simultaneously results in an accuracy score close to 0.500. To summarize, the optimal approach is highly dependent on the specific model and dataset, as demonstrated by the contradictory results on PrimeVul.

\paragraph{\bf Quality of Vulnerability Detection Datasets} 
\citet{ding2024vulnerability} recently highlighted key challenges in existing vulnerability detection datasets, including label noise, data duplication, and data leakage. To mitigate these concerns, we incorporate the PrimeVul dataset in our experiments, as it was carefully curated to address such issues, along with several other commonly used datasets. We also include an additional unpublished dataset (NodeMedic) to demonstrate the generality of our approach. Beyond the noisiness of current vulnerability datasets, we acknowledge that function-level vulnerability detection has inherent limitations compared to repository-level detection~\cite{risse2025top}. However, we believe that function-level datasets and detectors provide a valuable first step toward addressing vulnerability detection in broader contexts.

\section{Conclusion}

This work studies the effectiveness of different few-shot selection methods for large language models in code vulnerability detection. We evaluated several common techniques and introduced combined methods built upon them, showing that while open-source models perform worse under baseline few-shot settings, they achieve substantially greater improvements with our combined methods and can, in some cases, approach the performance of the closed-source model GPT-5-Mini.

\section*{Acknowledgment}

This work was supported in part by the National Science Foundation via grants No. 2504353 and 2504354, and by the Future Enterprise Security Initiative at Carnegie Mellon CyLab (FutureEnterprise@CyLab).

\bibliographystyle{IEEEtran}
\bibliography{Reference, Reference_2}

\appendices

\section{Full Results with Standard Deviation}
\begin{table*}[!htbp]
\centering
\makebox[\textwidth][c]{
\rotatebox{90}{
\begin{minipage}{\dimexpr\textheight-1.5\baselineskip\relax}
\caption{Full results with standard deviation for all approaches across three models. Each cell reports the mean and standard deviation over five runs for Gemma and Qwen models, while GPT results are based on a single run and do not have standard deviation. For the approach names, the following abbreviations are used: ZS = zero-shot, R-FS = random few-shot, LFM = Learn-from-Mistakes, LFNN = Learn-from-Nearest-Neighbors, CM = Combined Method. A dash (‘–’) indicates that the metric is undefined in at least one of the runs due to division by zero.
}
\label{table:main_full}
\centering
\resizebox{!}{0.26\textheight}{
\begin{tabular}{llrrrrrrrrrrrr}
\toprule
Dataset & Approach & \multicolumn{4}{c}{Gemma-3-4b-it} & \multicolumn{4}{c}{GPT-5-mini} & \multicolumn{4}{c}{Qwen-2.5-Coder} \\
\cmidrule(lr){3-6} \cmidrule(lr){7-10} \cmidrule(lr){11-14}
& & Acc & Prec & Recall & F1 & Acc & Prec & Recall & F1 & Acc & Prec & Recall & F1 \\
\midrule
\multirow{7}{*}{\rotatebox{90}{DiverseVul}} & ZS & 0.619 (±0.003) & 0.770 (±0.009) & 0.340 (±0.000) & 0.472 (±0.002)  & 0.593 & 0.588 & 0.627 & 0.607 & 0.497 (±0.000) & 0.000 (±0.000) & 0.000 (±0.000) & 0.000 (±0.000)  \\
& R-FS & 0.545 (±0.053) & 0.571 (±0.081) & 0.467 (±0.210) & 0.480 (±0.132)  & 0.560 & 0.598 & 0.367 & 0.455 & 0.599 (±0.045) & 0.741 (±0.027) & 0.303 (±0.139) & 0.411 (±0.150)  \\
& LFM & 0.525 (±0.003) & 0.840 (±0.056) & 0.061 (±0.005) & 0.114 (±0.009)  & 0.617 & 0.647 & 0.513 & 0.573 & 0.500 (±0.000) & 0.500 (±0.000) & 1.000 (±0.000) & 0.667 (±0.000)  \\
& LFNN & 0.512 (±0.004) & 0.511 (±0.004) & 0.537 (±0.009) & 0.524 (±0.006)  & 0.590 & 0.624 & 0.453 & 0.525 & 0.659 (±0.009) & 0.632 (±0.007) & 0.759 (±0.014) & 0.690 (±0.009)  \\
& CM 1 & 0.548 (±0.020) & 0.530 (±0.015) & 0.868 (±0.035) & 0.658 (±0.007)  & 0.587 & 0.610 & 0.480 & 0.537 & 0.638 (±0.008) & 0.606 (±0.011) & 0.792 (±0.039) & 0.686 (±0.011)  \\
& CM 2 & 0.515 (±0.002) & 0.508 (±0.001) & 0.973 (±0.000) & 0.667 (±0.001)  & 0.560 & 0.573 & 0.473 & 0.518 & 0.597 (±0.006) & 0.557 (±0.003) & 0.949 (±0.008) & 0.702 (±0.004)  \\
& CM 3 & 0.499 (±0.002) & 0.499 (±0.001) & 0.985 (±0.007) & 0.663 (±0.002)  & 0.570 & 0.585 & 0.480 & 0.527 & 0.531 (±0.021) & 0.517 (±0.012) & 0.988 (±0.015) & 0.678 (±0.007)  \\
\midrule
\multirow{7}{*}{\rotatebox{90}{NodeMedic}} & ZS & 0.506 (±0.003) & 0.765 (±0.005) & 0.460 (±0.000) & 0.574 (±0.001)  & 0.757 & 0.786 & 0.912 & 0.845 & 0.375 (±0.009) & 0.870 (±0.049) & 0.162 (±0.005) & 0.273 (±0.008)  \\
& R-FS & 0.687 (±0.060) & 0.726 (±0.003) & 0.914 (±0.136) & 0.804 (±0.060)  & 0.794 & 0.845 & 0.876 & 0.860 & 0.632 (±0.060) & 0.748 (±0.020) & 0.750 (±0.163) & 0.739 (±0.071)  \\
& LFM & 0.725 (±0.000) & 0.725 (±0.000) & 1.000 (±0.000) & 0.840 (±0.000)  & 0.767 & 0.789 & 0.927 & 0.852 & 0.720 (±0.000) & 0.723 (±0.000) & 0.993 (±0.000) & 0.837 (±0.000)  \\
& LFNN & 0.758 (±0.004) & 0.759 (±0.003) & 0.975 (±0.004) & 0.854 (±0.002)  & 0.788 & 0.849 & 0.861 & 0.855 & 0.713 (±0.011) & 0.768 (±0.004) & 0.866 (±0.014) & 0.814 (±0.008)  \\
& CM 1 & 0.701 (±0.005) & 0.755 (±0.009) & 0.870 (±0.018) & 0.808 (±0.004)  & 0.788 & 0.839 & 0.876 & 0.857 & 0.751 (±0.008) & 0.776 (±0.004) & 0.923 (±0.020) & 0.843 (±0.007)  \\
& CM 2 & 0.565 (±0.008) & 0.719 (±0.005) & 0.657 (±0.008) & 0.686 (±0.006)  & 0.794 & 0.840 & 0.883 & 0.861 & 0.735 (±0.006) & 0.748 (±0.004) & 0.958 (±0.003) & 0.840 (±0.003)  \\
& CM 3 & 0.722 (±0.007) & 0.728 (±0.003) & 0.984 (±0.014) & 0.837 (±0.005)  & 0.794 & 0.846 & 0.883 & 0.864 & 0.720 (±0.000) & 0.724 (±0.001) & 0.991 (±0.003) & 0.837 (±0.000)  \\
\midrule
\multirow{7}{*}{\rotatebox{90}{PrimeVul}} & ZS & 0.599 (±0.044) & 0.723 (±0.106) & 0.358 (±0.036) & 0.472 (±0.003) & 0.535 & 0.524 & 0.750 & 0.617 & 0.500 (±0.000) & -- & 0.002 (±0.004) & 0.004 (±0.008)  \\
& R-FS & 0.522 (±0.041) & 0.553 (±0.086) & 0.373 (±0.148) & 0.422 (±0.101) & 0.560 & 0.556 & 0.600 & 0.577 & 0.503 (±0.010) & 0.528 (±0.058) & 0.232 (±0.105) & 0.304 (±0.094)  \\
& LFM & 0.516 (±0.013) & 0.726 (±0.186) & 0.437 (±0.459) & 0.336 (±0.270) & 0.550 & 0.557 & 0.490 & 0.521 & 0.500 (±0.000) & 0.500 (±0.000) & 1.000 (±0.000) & 0.667 (±0.000)  \\
& LFNN & 0.527 (±0.030) & 0.520 (±0.017) & 0.648 (±0.147) & 0.571 (±0.064) & 0.555 & 0.549 & 0.620 & 0.582 & 0.483 (±0.009) & 0.390 (±0.056) & 0.060 (±0.009) & 0.104 (±0.015)  \\
& CM 1 & 0.543 (±0.032) & 0.527 (±0.022) & 0.885 (±0.040) & 0.659 (±0.010) & 0.575 & 0.577 & 0.600 & 0.588 & 0.473 (±0.007) & 0.462 (±0.010) & 0.336 (±0.092) & 0.383 (±0.057)  \\
& CM 2 & 0.516 (±0.001) & 0.508 (±0.001) & 0.964 (±0.022) & 0.666 (±0.005) & 0.555 & 0.549 & 0.610 & 0.578 & 0.455 (±0.012) & 0.445 (±0.015) & 0.370 (±0.026) & 0.404 (±0.021)  \\
& CM 3 & 0.500 (±0.002) & 0.500 (±0.001) & 0.982 (±0.017) & 0.663 (±0.004) & 0.555 & 0.546 & 0.650 & 0.594 & 0.499 (±0.006) & 0.499 (±0.003) & 0.986 (±0.015) & 0.663 (±0.005)  \\
\midrule
\multirow{7}{*}{\rotatebox{90}{SVENC}} & ZS & 0.478 (±0.000) & 0.474 (±0.000) & 0.400 (±0.000) & 0.434 (±0.000)  & 0.533 & 0.523 & 0.756 & 0.618 & 0.500 (±0.000) & -- & 0.000 (±0.000) & 0.000 (±0.000)  \\
& R-FS & 0.506 (±0.010) & 0.503 (±0.005) & 0.778 (±0.347) & 0.564 (±0.174)  & 0.600 & 0.579 & 0.733 & 0.647 & 0.491 (±0.013) & -- & 0.422 (±0.373) & 0.345 (±0.274)  \\
& LFM & 0.500 (±0.000) & 0.500 (±0.000) & 1.000 (±0.000) & 0.667 (±0.000)  & 0.544 & 0.543 & 0.556 & 0.549 & 0.500 (±0.000) & 0.500 (±0.000) & 1.000 (±0.000) & 0.667 (±0.000)  \\
& LFNN & 0.471 (±0.009) & 0.428 (±0.034) & 0.187 (±0.044) & 0.258 (±0.049)  & 0.567 & 0.554 & 0.689 & 0.614 & 0.504 (±0.009) & 0.567 (±0.133) & 0.040 (±0.009) & 0.075 (±0.017)  \\
& CM 1 & 0.520 (±0.008) & 0.513 (±0.005) & 0.813 (±0.027) & 0.629 (±0.010)  & 0.578 & 0.566 & 0.667 & 0.612 & 0.520 (±0.019) & 0.527 (±0.024) & 0.387 (±0.074) & 0.442 (±0.055)  \\
& CM 2 & 0.536 (±0.004) & 0.520 (±0.003) & 0.947 (±0.011) & 0.671 (±0.002)  & 0.589 & 0.574 & 0.689 & 0.626 & 0.493 (±0.005) & 0.495 (±0.004) & 0.636 (±0.011) & 0.556 (±0.005)  \\
& CM 3 & 0.500 (±0.000) & 0.500 (±0.000) & 0.938 (±0.029) & 0.652 (±0.007)  & 0.611 & 0.596 & 0.689 & 0.639 & 0.496 (±0.005) & 0.498 (±0.003) & 0.982 (±0.009) & 0.661 (±0.004)  \\
\midrule
\multirow{7}{*}{\rotatebox{90}{SVENP}} & ZS & 0.587 (±0.006) & 0.597 (±0.007) & 0.537 (±0.013) & 0.565 (±0.009)  & 0.763 & 0.727 & 0.842 & 0.780 & 0.705 (±0.011) & 0.965 (±0.028) & 0.426 (±0.020) & 0.591 (±0.019)  \\
& R-FS & 0.558 (±0.038) & 0.561 (±0.044) & 0.632 (±0.117) & 0.584 (±0.035)  & 0.789 & 0.806 & 0.763 & 0.784 & 0.616 (±0.027) & 0.707 (±0.071) & 0.432 (±0.144) & 0.514 (±0.102)  \\
& LFM & 0.500 (±0.000) & 0.500 (±0.000) & 0.905 (±0.013) & 0.644 (±0.003)  & 0.776 & 0.744 & 0.842 & 0.790 & 0.500 (±0.000) & 0.500 (±0.000) & 1.000 (±0.000) & 0.667 (±0.000)  \\
& LFNN & 0.634 (±0.005) & 0.641 (±0.004) & 0.611 (±0.011) & 0.625 (±0.007)  & 0.895 & 0.941 & 0.842 & 0.889 & 0.792 (±0.005) & 0.845 (±0.002) & 0.716 (±0.011) & 0.775 (±0.007)  \\
& CM 1 & 0.553 (±0.012) & 0.546 (±0.009) & 0.616 (±0.043) & 0.579 (±0.023)  & 0.882 & 0.892 & 0.868 & 0.880 & 0.787 (±0.023) & 0.793 (±0.034) & 0.779 (±0.021) & 0.785 (±0.019)  \\
& CM 2 & 0.632 (±0.000) & 0.639 (±0.000) & 0.605 (±0.000) & 0.622 (±0.000)  & 0.895 & 0.917 & 0.868 & 0.892 & 0.697 (±0.019) & 0.660 (±0.016) & 0.816 (±0.029) & 0.729 (±0.018)  \\
& CM 3 & 0.521 (±0.013) & 0.512 (±0.008) & 0.942 (±0.045) & 0.663 (±0.009)  & 0.803 & 0.795 & 0.816 & 0.805 & 0.600 (±0.018) & 0.558 (±0.012) & 0.963 (±0.027) & 0.707 (±0.010)  \\
\bottomrule
\end{tabular}
}
\end{minipage}
}
}
\end{table*}

An extended version of Table~\ref{table:main_no_std} with standard deviation is shown in Table~\ref{table:main_full}.
A key observation is the general consistency of the methods---most exhibit low standard deviations. Table ~\ref{table:main_full} reinforces the reliability of the mean results discussed in the main text.

As anticipated, the R-FS approach consistently displayed the highest variance, underscoring its sensitivity to the random selection of examples. This is particularly evident on datasets like SVENC with the Qwen model, where the Recall was 0.422 (±0.373). In contrast, the failure cases of certain baselines and methods proved to be remarkably deterministic. For instance, the LFM method's tendency to uniformly predict the positive class on datasets like NodeMedic is confirmed by its perfect recall with zero deviation (1.000 ±0.000) for both Gemma and Qwen models.

Furthermore, the stability of our more successful methods lends additional credence to their effectiveness. The LFNN approach on NodeMedic not only achieved a high F1-score (0.854 for Gemma) but did so with minimal variance (±0.002). Similarly, the Combined Methods on the DiverseVul dataset demonstrated lower variance than R-FS, suggesting that their improved performance is not an artifact of random chance.

\end{document}